\documentclass[showpacs,preprintnumbers,prd,amsmath,amssymb]{revtex4}
\usepackage{amsmath}
\usepackage{graphicx}
\usepackage{subfig}
\usepackage{hyperref}
\usepackage{amssymb}
\usepackage[english]{babel}
\usepackage{epsfig}
\usepackage{wasysym}
\usepackage{color,xcolor}

\begin{document}

\title{Bardeen Regular Black Hole With an Electric Source\\}

\author{ Manuel E. Rodrigues$^{(a,b)}$\footnote{E-mail address: 
esialg@gmail.com}, Marcos V. de S. Silva$^{(b)}$\footnote{E-mail 
address: marco2s303@gmail.com}}
\affiliation{$^{(a)}$Faculdade de Ci\^{e}ncias Exatas e Tecnologia, 
Universidade Federal do Par\'{a}\\
Campus Universit\'{a}rio de Abaetetuba, 68440-000, Abaetetuba, Par\'{a}, 
Brazil\\
$^{(b)}$Faculdade de F\'{\i}sica, Programa de P\'os-Gradua\c{c}\~ao em 
F\'isica, Universidade Federal do 
 Par\'{a}, 66075-110, Bel\'{e}m, Par\'{a}, Brazil}


\begin{abstract}
If some energy conditions on the stress-energy tensor are violated, is possible construct regular black holes in General Relativity and in alternative theories of gravity. This type of solution has horizons but does not present singularities. The first regular black hole was presented by Bardeen and can be obtained from Einstein equations in the presence of an electromagnetic field. E. Ayon-Beato and A. Garcia reinterpreted the Bardeen metric as a magnetic solution of General Relativity coupled to a nonlinear electrodynamics. In this work show that the Bardeen model may also be interpreted as a solutions of Einstein equations in the presence of a electric source, whose electric field does not behaves as a Coulomb field. We analyzed the asymptotic forms of the Lagrangian for the electric case and also analyzed the energy conditions. 
\end{abstract}

\date{\today}

\maketitle



\section{Introduction}
\label{sec1}
In 1916 Einstein proposed a relativistic theory for gravitation field. This theory, know as General Relativity, describe the gravitational interaction as a result of the curvature of the spacetime, which is generated for the presence of matter and energy in that spacetime \cite{din}. This physics are mathematically synthesized by the Einstein equations \cite{wal}. Besides explained phenomenons that are not consistent with Newton's theory, as the precession of the perihelion of Mercury \cite{will1}, the Einstein theory predicted new facts as the bending of light, measured by Eddington in 1919 \cite{will2}, and the gravitational waves, detected by the LIGO/Virgo Collaboration \cite{Ligo1,Ligo2,Ligo3,Ligo4}.

One of the most interested predictions of General Relativity are black holes. These astrophysical objects, that are solutions of Einstein equations \cite{chan}, are a great candidate to test General Relativity and modified theories of gravity due to the strong gravitational field. In general Relativity, by the no-hair theorem, black holes can be characterized for three parameters: mass, charge and angular momentum \cite{BCH,Grav}. The most simple example is a black hole characterized only by his mass, Schwarzschild black \cite{las}. There are other solutions more general than Schwarzschild, as Reissner-Nordström (black hole with mass and electric charge \cite{chan,las}) and Kerr (black hole with mass and angular momentum \cite{chan,vis1,kerr}).

A way to extract information about black holes is analyzing the behavior of fields and particles around them \cite{vitor1,Herdeiro1,Herdeiro2,Herdeiro3,Herdeiro4}. In this sense, is interesting study absorption, scattering and quasinormal modes of fields with different spins from different types of black holes. Is also important try to understand the inner structure of black holes. Actually is inside the trapped surface (event horizon) that are one of the biggest problem, the presence of singularities. Singularities could be understood as a point where the geodesics are interrupted \cite{Bronnikov}. From the works of Hawking and Penrose is know that, if some energy conditions are satisfied by the stress-energy tensor, the presence of singularities are inevitably from a gravitational collapse \cite{Hawking1,Hawking2,Hawking3,Penrose}. The cosmic censorship conjecture says that this point, where the laws of physics lose the sense, must be hidden by a event horizon, protecting the exterior spacetime.

A way to avoid the singularity was proposed by Gliner and Sakarov, where the matter source has a de sitter core with a equation of state $\rho=-p$ at the center of the spacetime \cite{Sakharov,Gliner}. After these works, James Bardeen proposed the first nonsingular solution of Einstein equations, the Bardeen regular black hole \cite{Bardeen}. This solution is regular in all spacetime, presenting a de sitter center, as suggest by the Sakharov's work, and satisfying the weak energy condition. As the Bardeen solution presented non-vanishing Einstein equations, Ayon-Beato and Garcia proposed that this metric could be interpreted as a solution of Einstein equations coupled with a nonlinear Electrodynamics with a magnetic charge \cite{Beato1}.

Further solutions of regular black holes were found considering both magnetic and electric sources \cite{Beato2,Beato3,Beato4,Kirill1,Kirill12,Irina,Stefano,Leonardo2,balart,Nami,Ponce,Manuel1}, some cases with rotation \cite{bambi,neves,toshmatov,azreg,DYM,ramon} and others in alternative theories os gravity \cite{berej,rodrigues1,rodrigues2,rodrigues3}. Actually for each solution with an electric source is possible construct the same solution with a magnetic source \cite{Kirill1,Irina}. In this sense, is interesting see if is possible reconstruct the Bardeen regular black hole as an electrically charged solution of Einstein equations, which is the main objective of this paper.

The structure os this work is organized as follows. In Sec. \ref{sec2} we make a brief review of the Bardeen regular black hole with magnetic monopole. In Sec. \ref{sec3} we will take the Bardeen solution and construct a electric Lagrangian that describes this system. In Sec. \ref{sec4} we will analyze which energy conditions this solution satisfy, theses conditions must be the same for the magnetic and electric cases. Our conclusions and final remarks are present in Sec. \ref{sec5}. In this work we are considering the metric signature ($+,-,-,-$) and natural units, where $c=\hbar=G=1$. 
\section{Regular black hole with magnetic source}
\label{sec2}
The first solution of Einstein equations that describe black holes without singularities was proposed by James Bardeen in 1968. The physical interpretation of the Bardeen metric was shown by Ayon-Beato and Garcia. This solution can describe black holes with a nonlinear magnetic monopole resulting in a solution of Einstein equations coupled to a nonlinear electrodynamics. General relativity within nonlinear electrodynamics can be described by the action
\begin{equation}	
S=\int d^4x \sqrt{-g}\left[R+2\kappa^2 \mathcal{L}(F)\right],
\label{BAC}
\end{equation}
where $R$ is de curvature scalar and $\mathcal{L}(F)$ is the Lagrangian Density of the electromagnetic field, with $F=\frac{1}{4}F^{\mu\nu}F_{\mu\nu}$, where $F_{\mu\nu}$ is the Faraday-Maxwell tensor, and $\kappa^2=8\pi$. Varying the action \eqref{BAC} with respect to the metric $g_{\mu\nu}$ we get the Einstein equations, given by
\begin{equation}
R_{\mu\nu}-\frac{1}{2}g_{\mu\nu}=\kappa^2 T_{\mu\nu},
\label{EE}
\end{equation}
where $R_{\mu\nu}$ is the Ricci tensor and $T_{\mu\nu}$ is the stress-energy tensor. The stress-energy tensor can be write as
\begin{equation}
T_{\mu\nu}=g_{\mu\nu}\mathcal{L}-\frac{\partial\mathcal{L}(F)}{\partial F}F_{\mu}^{\ \alpha}F_{\alpha\nu}.
\label{ST}
\end{equation}

As the Faraday-Maxwell tensor is given in terms of a gauge potential, $A_\mu$, in the form $F_{\mu\nu}=\partial_\mu A_\nu-\partial_\nu A_\mu$, we can obtain the Maxwell equations for a nonlinear electrodynamics varying the action \eqref{BAC} with respect to $A_\mu$. These equations are given by
\begin{equation}
\nabla_\mu\left[F^{\mu\nu}\mathcal{L}_F \right]\equiv \partial_\mu\left[\sqrt{-g}F^{\mu\nu}\mathcal{L}_F\right],
\label{ME}
\end{equation}
where $\mathcal{L}_F=\partial\mathcal{L}/\partial F$.

To obtain the Bardeen metric, Ayon-Beato and Garcia used a Lagrangian density written as 
\begin{equation}
\mathcal{L}(F)=\frac{3}{sq^2_{BD}\kappa^2}\left(\frac{\sqrt{2q^2_{BD}F}}{1+\sqrt{2q^2_{BD}F}}\right)^{5/2},
\end{equation}
with $s=\left.\right|q_{BD} \left|\right. / (2m)$, $q_{BD}$ is the magnetic charge and $m$ is the ADM mass,and a line element, that describe a spherically symmetric spacetime, as
\begin{equation}
ds^2=f(r)dt^2-f(r)^{-1}dr^2-r^2\left(d\theta^2+\sin^2\theta d\phi^2\right),
\label{le}
\end{equation}
where
\begin{equation}
f(r)=1-\frac{2M(r)}{r}.
\label{f}
\end{equation}

From the Maxwell equations, we can prove that the magnetic field has the form
\begin{equation}
F_{23}(\theta)=q_{BD}\sin\theta.
\end{equation}
Since we have the components of the Maxwell-Faraday tensor, we calculate the scalar $F$ as
\begin{equation}
F(r)=\frac{q_{BD}^2}{2r^4},
\end{equation}
and so that the Lagrangian density is written  in terms of the radial coordinate,
\begin{equation}
\mathcal{L}(r)=\frac{3}{sq_{BD}^2\kappa^2}\left(\frac{q_{BD}^2}{r^2+q_{BD}^2}\right)^{5/2}.
\label{LM}
\end{equation}

The asymptotic forms for $F\rightarrow +\infty$ ($r\rightarrow 0$) and $F\rightarrow 0$ ($r\rightarrow +\infty$) are
\begin{eqnarray}
&&\mathcal{L}(F)\approx \frac{3}{sq_{BD}^2\kappa^2}-\frac{15m}{\kappa^2\sqrt{2F}q_{BD}^4},\ \mbox{for}\ F\rightarrow +\infty,\\
&&\mathcal{L}(F)\approx \frac{6m\left(2F\right)^{5/4}}{\sqrt{\left|q_{BD}\right|}\kappa^2},\ \mbox{for}\ F\rightarrow 0.
\end{eqnarray}
So, for small values of $F$ de Lagrangian becomes zero and for great values the Lagrangian becomes a constant.

\begin{figure}
	\includegraphics[scale=0.6]{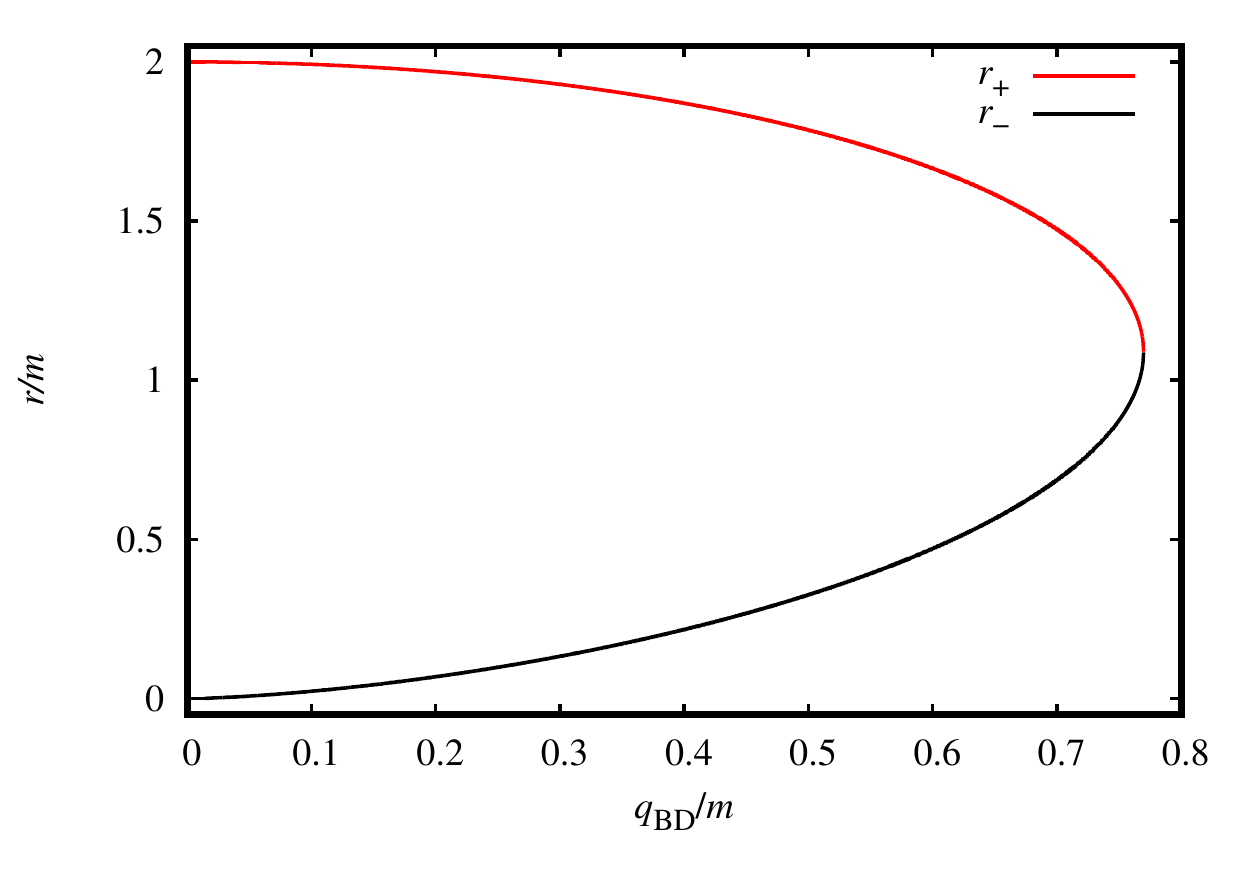}
	\caption{Event horizon and Cauchy horizon of Bardeen black hole as a function of the charge.}
	\label{hr}
\end{figure}

For the line element \eqref{le}, the non-zero and independents components of Einstein equations are
\begin{eqnarray}
\frac{2 M'(r)}{r^2}&=&\kappa^2\mathcal{L}(F),\label{eq1}\\
\frac{M''(r)}{r}&=&\kappa^2\left[\mathcal{L}-\mathcal{L}_F F^{23}F_{23}\right].\label{eq2}
\end{eqnarray}
As we have $\mathcal{L}(F)$ in terms of the radial coordinate, from \eqref{eq1} we can write
\begin{equation}
M'(r)=\frac{3mq_{BD}^2r^2}{\left(r^2+q_{BD}^2\right)^{5/2}}.
\label{DM}
\end{equation}
Integrating \eqref{DM} we obtain the mass function that generates the Bardeen regular black hole,
\begin{equation}
M(r)=\frac{mr^3}{\left(r^2+q_{BD}^2\right)^{3/2}}.
\label{M}
\end{equation}
The horizons associated with this solution can be find calculating $f(r)=0$. In Fig. \ref{hr} we show the behavior of the event horizon and Cauchy horizon in relation of the charge. We can see that de radius of the event horizon decreases and the Cauchy horizon increases as the charge increases. The two horizons become one when $q_{BD}=4m/(3\sqrt{3})$, this configuration is known as extremal black hole. 
We can expand $f(r)$ for $r=0$ and $r\rightarrow+\infty$ to analyze the asymptotic forms. For $r\rightarrow+\infty$ we find
\begin{equation}
f(r)\approx 1-\frac{2m}{r}+\frac{3mq_{BD}^2}{r^3}.
\end{equation}
From this result we show that the Bardeen solution is not asymptotically Reissner-Nordström, however for points far enough the solution behaves as Schwarzschild and is asymptotically flat. For small values $r$ we have
\begin{equation}
f(r)\approx 1-\frac{2mr^2}{q_{BD}^3}.
\end{equation}
So, that Bardeen metric present a region inside the black hole that behaves as a de Sitter solution.
\section{Black holes electrically charged}\label{sec3}
\begin{figure}
	\includegraphics[scale=0.6]{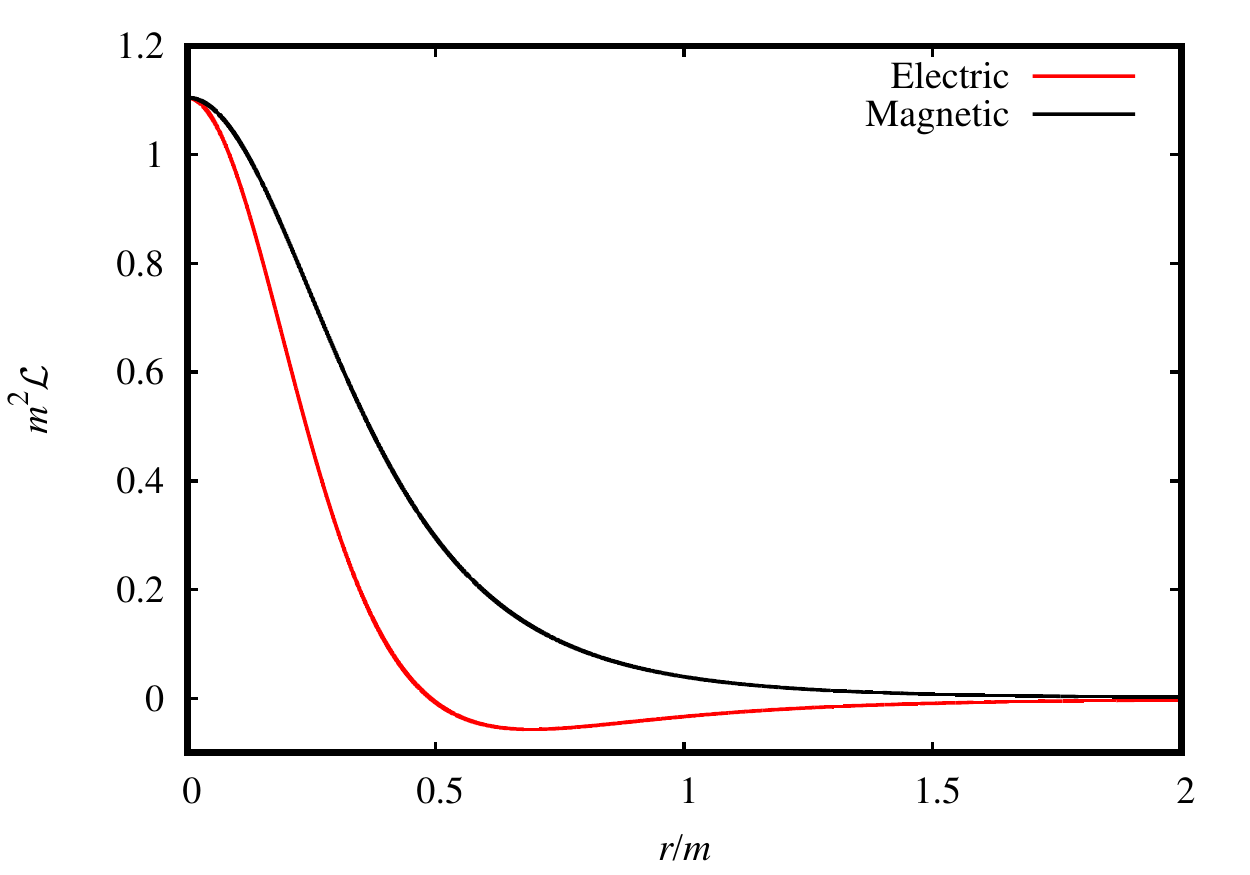}
	\caption{Behavior of $\mathcal{L}$ for the magnetic and electric case with $q_{BD}=0.6$.}
	\label{ELM}
\end{figure}

Using Einstein and Maxwell equations is possible associate the Bardeen black hole with an electric source. As the black hole is static and spherically symmetric, the only non-zero component of Maxwell-Faraday tensor is $F^{10}$. Integrating \eqref{ME} for $\nu=0$, we find
\begin{equation}
F^{10}(r)=\frac{q_{BD}}{r^2}\mathcal{L}_F^{-1}(r).
\label{F10}
\end{equation}
So that, the intensity of the electric field will be found since we have $\mathcal{L}_F$. For the electric case, we can write the non-zero components of Einstein equations as
\begin{eqnarray}
\frac{2 M'(r)}{r^2}&=&\kappa^2\left[\mathcal{L}+\frac{q^2_{BD}}{r^4}\mathcal{L}_F^{-1}\right],\label{eq01}\\
\frac{M''(r)}{r}&=&\kappa^2\mathcal{L}\label{eq02}.
\end{eqnarray}
We may solve equations \eqref{eq01}-\eqref{eq02} for $\mathcal{L}$ and $\mathcal{L}_F$ to get
\begin{eqnarray}
\mathcal{L}(r)&=&\frac{q^2_{BD}m\left(6q^2_{BD}-9r^2\right)}{\kappa^2\left(r^2+q^2_{BD}\right)^{7/2}},\label{Lang}\\
\mathcal{L}_{F}(r)&=&\frac{\kappa^2\left(r^2+q^2_{BD}\right)^{7/2}}{15mr^6}.\label{DevL}
\end{eqnarray}
In fig. \ref{ELM} we may compare the Lagrangian for the magnetic and electric interpretation. These functions are deferents, however for points closely the origin the functions have the same value and for the infinite of radial coordinate they tends to zero.

\begin{figure}
	\includegraphics[scale=0.6]{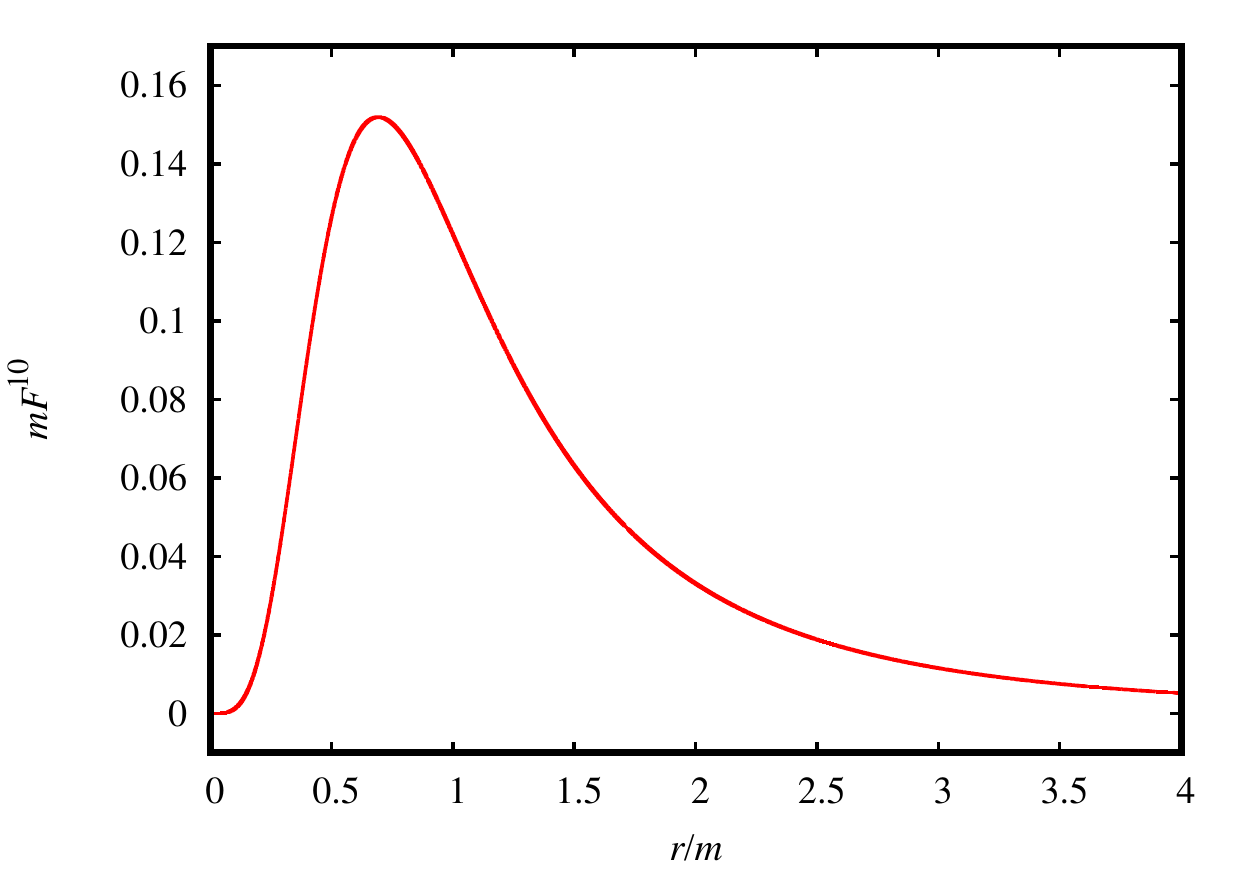}
	\caption{Intensity of electric field for Bardeen with $q_{BD}=0.6m$.}
	\label{f10}
\end{figure}

\begin{figure}
	\includegraphics[scale=0.6]{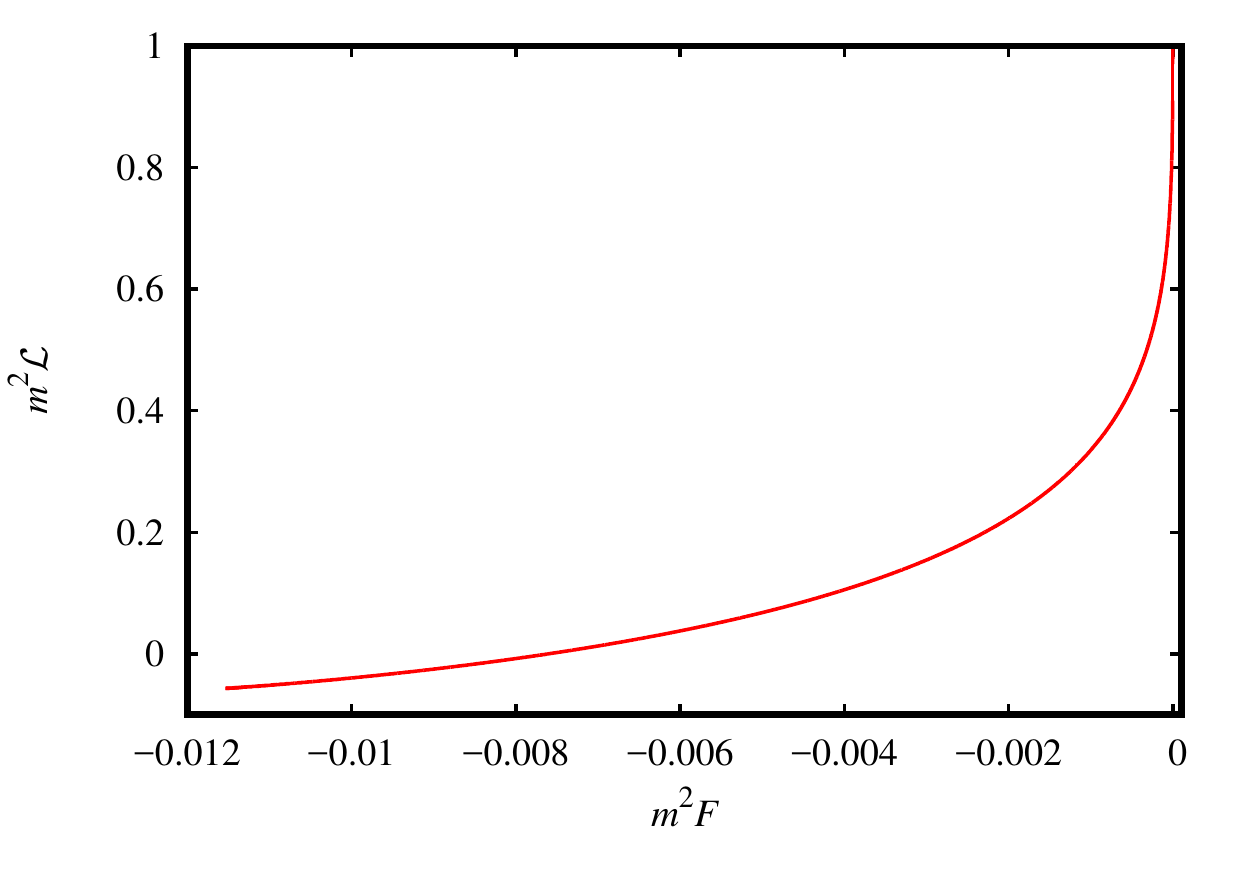}
	\caption{Behavior of the electric Lagrangian $\mathcal{L}(F)$ as a function of $F$ with $q=0.6m$.}
	\label{EL}
\end{figure}

Replacing \eqref{DevL} into \eqref{F10} we find
\begin{equation}
F^{10}(r)=\frac{15q_{BD}mr^4}{\kappa^2\left(r^2+q^2_{BD}\right)^{7/2}}.\label{CamB}
\end{equation}
In Fig. \ref{f10} we show the behavior of the electric field. We can see that the field is well behaved in the infinity and in the origin of radial coordinate and a maximum valor for
\begin{equation*}
r=\frac{2\left|q_{BD}\right|}{\sqrt{3}}.
\end{equation*}
To analyze the asymptotic forms we can expand \eqref{CamB} for $r\rightarrow 0$ and $r\rightarrow +\infty$, given by
\begin{eqnarray}
F^{10}(r)&\approx & \frac{15mq_{BD}}{\kappa^2 r^3}+O\left(\frac{1}{r^4}\right), \ \mbox{for} \ r\rightarrow +\infty.\\
F^{10}(r)&\approx & \frac{15mr^4 Sign(q_{BD})}{\kappa^2 q_{BD}^6}+O\left(r^5\right),\ \mbox{for} \ r \rightarrow 0.
\end{eqnarray}
With that we can see that this field do not behaves like a Coulomb field. The $F$ scalar becomes
\begin{equation}
F(r)=-\frac{225 m^2 q_{BD}^2 r^8}{2 \kappa ^4 \left(q_{BD}^2+r^2\right)^7}.
\label{F}
\end{equation}
\begin{figure}
	\includegraphics[scale=0.6]{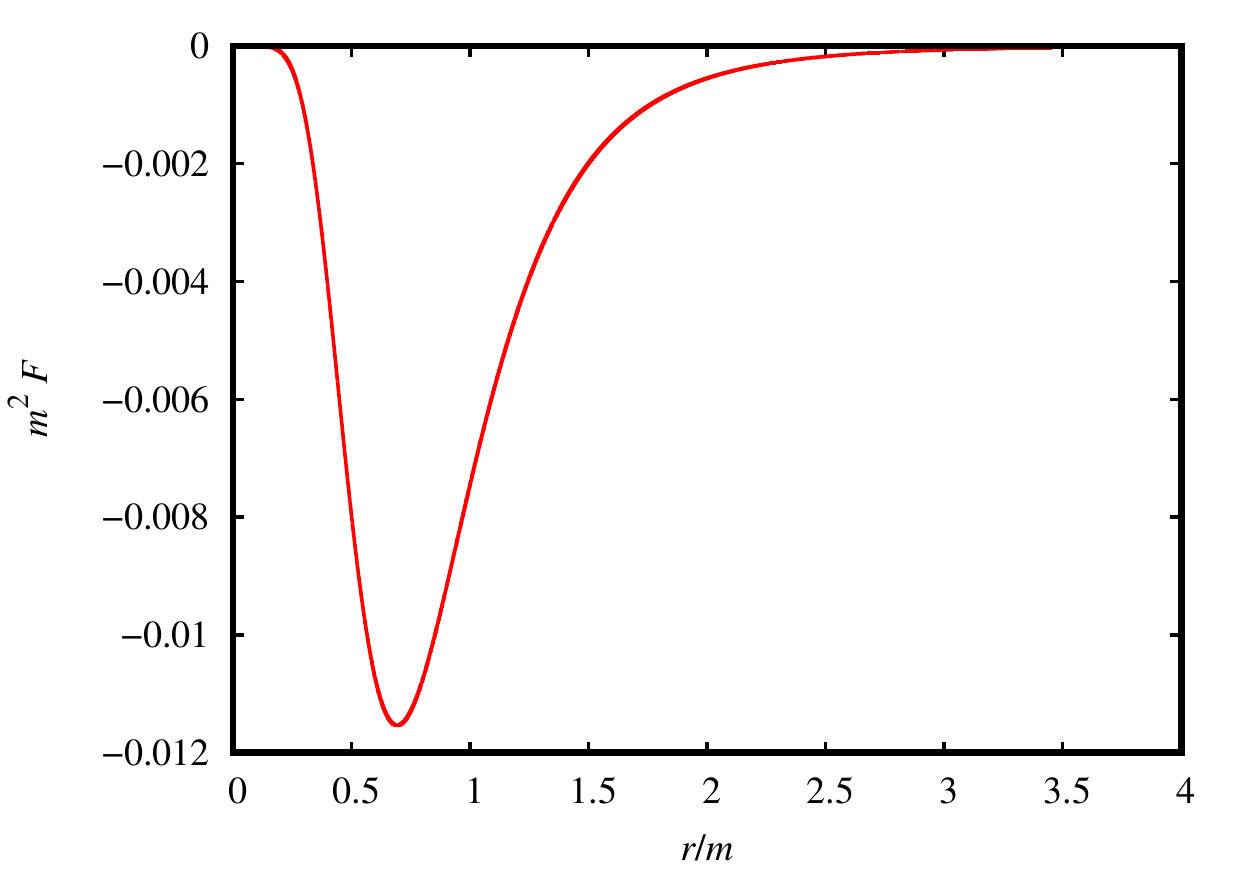}
	\caption{Behavior of the scalar $F$ as a function of the  radial coordinate for $q_{BD}=0.6m$.}
	\label{FF}
\end{figure}
From fig. \ref{F} we realize that the scalar $F(r)$ goes to zero for small and for big values of $r$. This function has a minimum value at the same point where $F^{10}(r)$ has a maximum. Now we can inverted equation \eqref{F} in order to write $r$ as a function of $F$ and obtain $\mathcal{L}(F)$. However, it's not possible write $\mathcal{L}(F)$ in a closed form, so that, we may analyzed the asymptotically forms or make a parametric plot showing the behavior of $\mathcal{L}(F)\times F$. In fig. \ref{EL} becomes clearly the nonlinear behavior of the electromagnetic theory. Analyzing the asymptotic cases we find that
\begin{eqnarray}
&&\mathcal{L}(F)\approx \frac{3}{sq_{BD}^2\kappa^2}-\frac{4 (-2F)^{1/4}\sqrt{30m\pi}}{q_{BD}^2\kappa^2},\ \mbox{for}\ r\rightarrow 0,\\
&&\mathcal{L}(F)\approx  \frac{ 2^{5/6} (-9F^5)^{1/6} \kappa ^{4/3} q_{BD}^2}{5\ (5m)^{2/3} \left| q_{BD}\right| ^{5/3}},\ \mbox{for}\ r\rightarrow +\infty.
\end{eqnarray} 

\begin{figure}
	\includegraphics[scale=0.6]{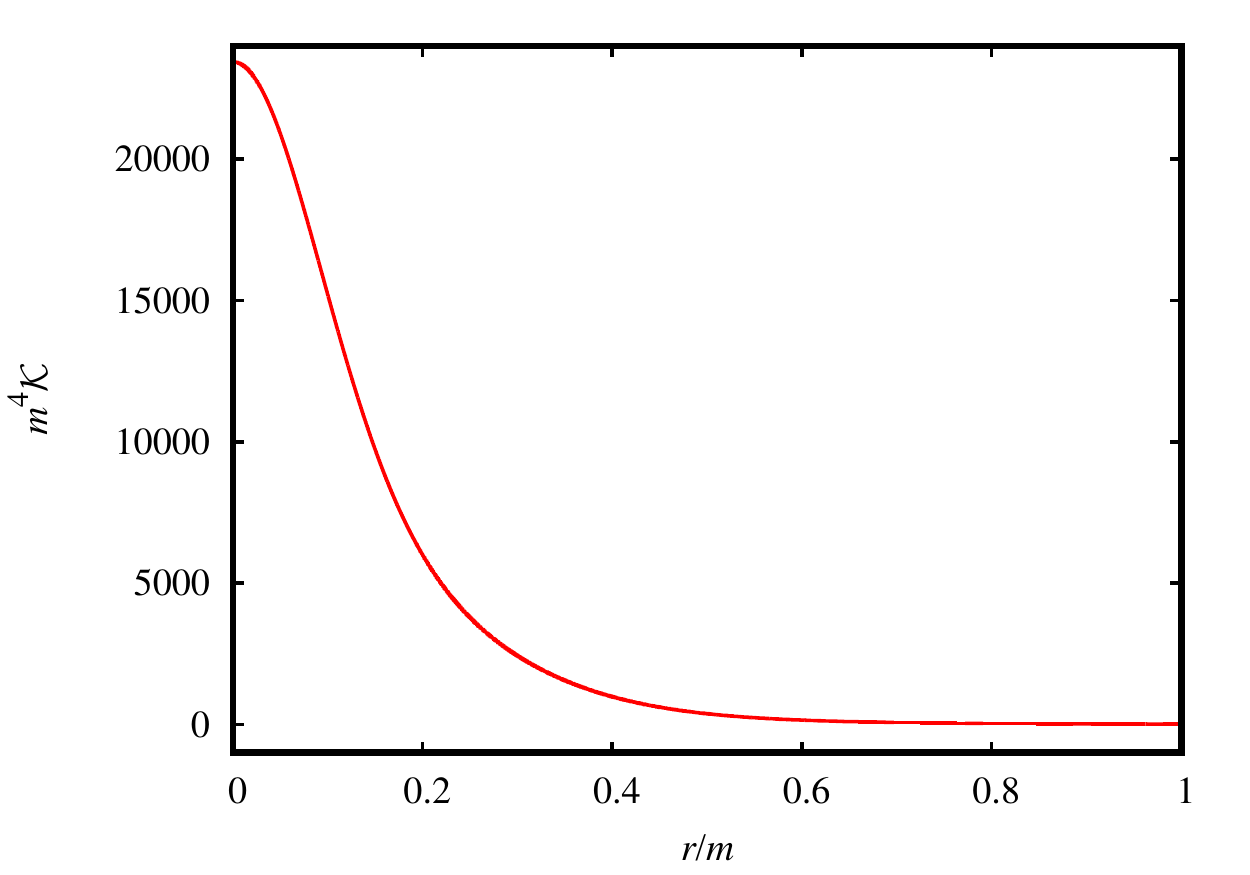}
	\caption{Kretschmann scalar for the Bardeen metric with $q_{BD}=0.6m$.}
	\label{Kr}
\end{figure}

Is also important analyze the regularity of the spacetime. In order to do that, we need verify if all curvature invariants are finite for all values do the radial coordinate. Actually, as we are working with a spherically symmetric ans static spacetime, if the Kretschmann scalar, $\mathcal{K}=R^{\mu\nu\alpha\beta}R_{\mu\nu\alpha\beta}$, is regular so all curvatures invariants will presented the same behavior. For the line element \eqref{le} with \eqref{f} and \eqref{M}, the Kretschmann scalar is given by:
\begin{eqnarray}
\mathcal{K}(r)&=&\frac{12m^2}{\left(r^2+q^2_{BD}\right)^7}\left(8q^8_{BD}-4q^6_{BD}r^2\right.+\left.47q^4_{BD}r^4-12q^2_{BD}r^6+4r^8\right).
\label{Kre}
\end{eqnarray}
In fig. \ref{Kr} we plot the equation \eqref{Kre} as a function of $r$ and we can see that this is well behaved in all spacetime. The solution is regular in te center, where the Kretschmann scalar is a constant, and in the infinity, where we have a flat spacetime with $\lim\limits_{r\rightarrow+\infty}\mathcal{K}(r)=0$.

\section{Energy conditions}\label{sec4}
In order to verify if the solution has physical sense, in other words if the solution represent a realistic source, we can verify the energy conditions. Performing the identifications $T^{0}_{\ 0}=\rho$, $T^{1}_{\ 1}=-p_r$ and $T^{2}_{\ 2}=T^{3}_{\ 3}=-p_t$, where $\rho$ is the energy density, $p_r$ is the radial pressure and $p_t$ the tangential pressure, the energy conditions are given by:
\begin{eqnarray}
SEC(r)&=&\rho+p_r+2p_t \geq 0,\label{ECG1}\\
WEC_{1,2}(r)&=&NEC_{1,2}(r)=\rho+p_{r,t}\geq 0,\label{ECG2}\\
WEC_{3}(r)&=&DEC_{1}(r)=\rho \geq 0,\label{ECG3}\\
DEC_{2,3}(r)&=&\rho-p_{r,t} \geq0.\label{ECG4}
\end{eqnarray}
Using the components of \eqref{ST}, we can write the energy density and the pressures as
\begin{eqnarray}
\rho(r)&=&\frac{6mq_{BD}^2}{\kappa^2\left(r^2+q_{BD}^2\right)^{5/2}},\label{rho}\\
p_r(r)&=&-\frac{6mq_{BD}^2}{\kappa^2\left(r^2+q_{BD}^2\right)^{5/2}},\label{pr}\\
p_t(r)&=&\frac{mq_{BD}^2\left(9r^2-6q_{BD}^2\right)}{\kappa^2\left(r^2+q_{BD}^2\right)^{7/2}}.\label{pt}
\end{eqnarray}
\begin{figure}
	\includegraphics[scale=0.6]{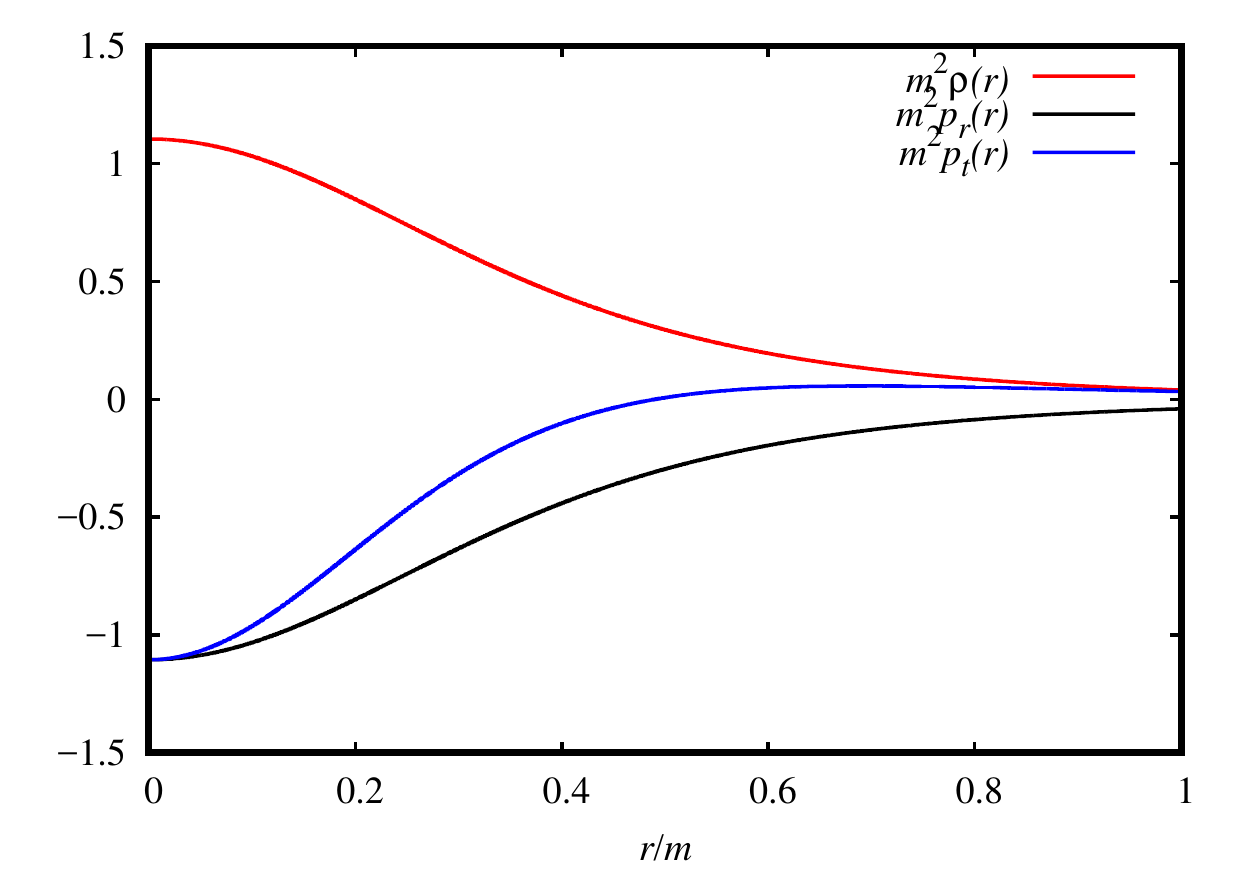}
	\caption{Energy density, radial pressure and tangential pressure as a function of the radial coordinate with $q_{BD}=0.6m$.}
	\label{Ener}
\end{figure}
In fig. \ref{Ener} we show the behavior of $\rho$, $p_r$ and $p_t$ in terms of $r$. The energy density is always positive and as \eqref{pr} is different of \eqref{pt}, we have a type of anisotropic perfect fluid with $p_r=-\rho$. At the center of the black hole $p_r\approx p_t$, resulting in a isotropic fluid with a de Sitter type equation os state $p\approx-\rho$.

\begin{figure}[!]
\begin{center}
	\includegraphics[scale=0.6]{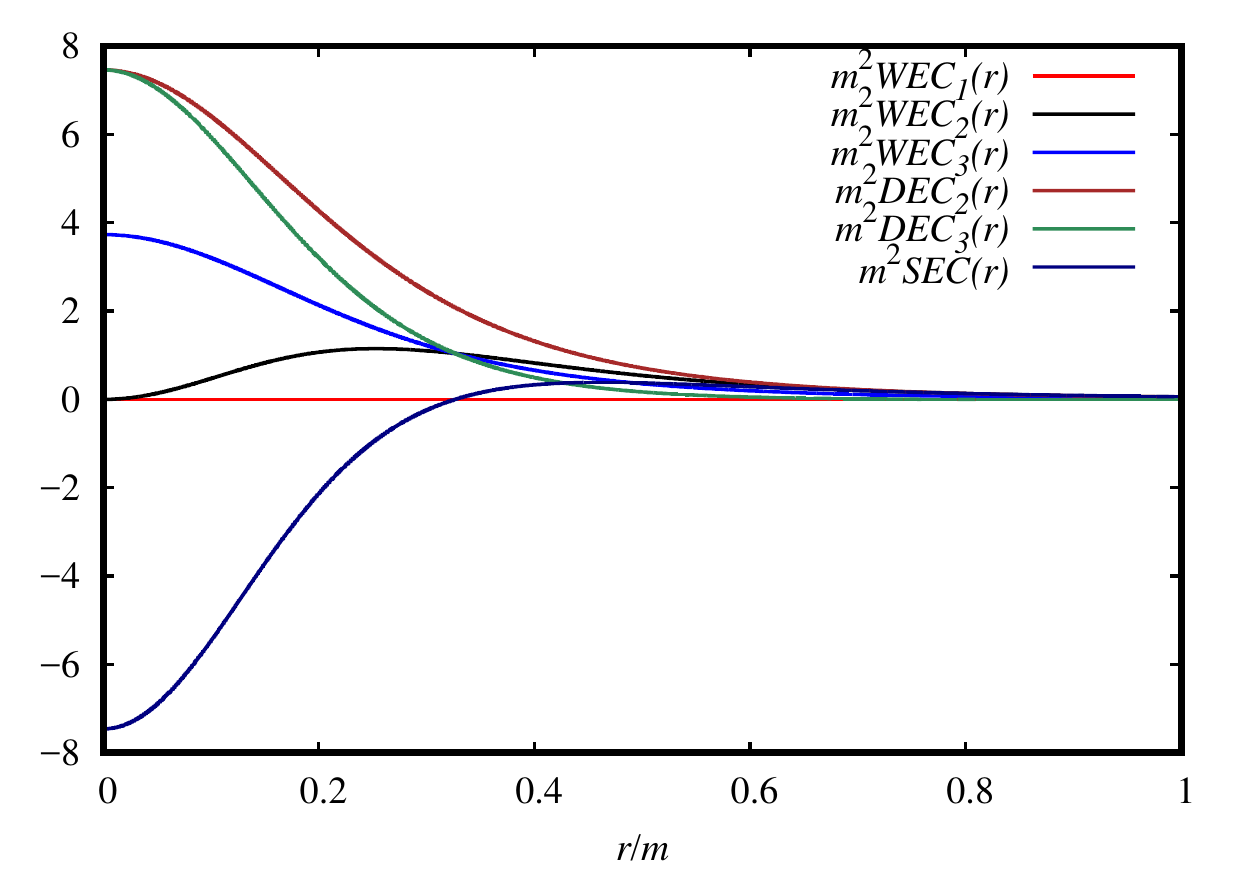}
	\caption{Graphical representations of the energy conditions for $q_{BD}=0.4m$.}
	\label{EC}
\end{center}
\end{figure}

Finally, the energy conditions are
\begin{eqnarray}
SEC(r)&=&\frac{6mq_{BD}^2\left(3r^2-2q_{BD}^2\right)}{\kappa^2\left(r^2+q_{BD}^2\right)^{7/2}},\\
WEC_{1}(r)&=&0,\\
WEC_{2}(r)&=&\frac{15mq_{BD}^2r^2}{\kappa^2\left(r^2+q_{BD}^2\right)^{7/2}},\\
WEC_{3}(r)&=&\frac{6mq_{BD}^2}{\kappa^2\left(r^2+q_{BD}^2\right)^{5/2}},\\
DEC_{2}(r)&=&\frac{12mq_{BD}^2}{\kappa^2\left(r^2+q_{BD}^2\right)^{5/2}},\\
DEC_{3}(r)&=&\frac{3mq_{BD}^2\left(4q_{BD}^2-r^2\right)}{\kappa^2\left(r^2+q_{BD}^2\right)^{7/2}}.
\end{eqnarray}
In fig. \ref{EC} we present the behave of the energy conditions varying with $r$ and we can see that for $r<\sqrt{2/3}\left|q_{BD}\right|$ the strong energy condition is violated. This region where SEC is violated is situated inside the event horizon as expected for black holes with a regular center.

\section{Conclusion}\label{sec5}
In this paper we review the interpretation of Bardeen solution, given by Ayon-Beato and Garcia, where we imposed a Lagrangian for a magnetic source and then, using Einstein equations, we obtained the mass function that generated the Bardeen metric. Despite the fact that the scalar $F$ diverge for $r\rightarrow 0$, the magnetic Lagrangian is well behaved for all points of the spacetime.

We also showed that the Bardeen model can be obtained for a electric source with a field that does not behaves as the Coulomb field but the scalar $F$ and Lagrangian are well behaved in all spacetime. As we can note write $\mathcal{L}(F)$ in a closed and analytical form, we make a parametric plot showing the nonlinear form in terms of $F$ and we also analyzed the asymptotic forms. For small values of $r$ the electric Lagrangian tends to a constant that has the same value for the magnetic case.

The Bardeen solution has two horizons, event horizon and Cauchy horizon. For some region inside the event horizon the strong energy condition is violated and presented a de Sitter center. The energy density is always positive and near $r=0$ a isotropic behaves appear. Through the Kretschmann scalar is possible see that the solution is regular in all spacetime and it's asymptotically flat. The energy conditions NEC and WEC are always satisfied, this result agree with the result obtained in \cite{Beato1}.  

It is important emphasize that for the electric interpretation $F$ vanish for $r\rightarrow 0$ and the electric field is zero too, in agreement with the Bronnikov's theorem \cite{Kirill1}. As $F$ vanish at the infinity and at the origin, this function has a minimum value at the same point where the electric field has a maximum. At least, the solution present a zero gravity surface, $\mathcal{L}=0$, that inner this surface the SEC is violated. These are the basics properties that are obligatorily for any electric Lagrangian according to \cite{Irina}. 

\vspace{1cm}

{\bf Acknowledgements}: M. E. R.  thanks Conselho Nacional de Desenvolvimento Cient\'ifico e Tecnol\'ogico - CNPq, Brazil, Edital MCTI/CNPQ/Universal 14/2014  for partial financial support.




\begin{thebibliography}{99}
%
\bibitem{din} R. D'Inverno, \textit{Introducing Einstein's Relativity}, Oxford University Press, New York (1998).

\bibitem{wal} R. M. Wald, \textit{General Relativity}, The University of Chicago Press, Chicago (1984).

\bibitem{will1} C. M. Will, \textit{Was Einstein Right? A Centenary Assessment}, \href{https://arxiv.org/abs/1409.7871}{{\tt arXiv:1409.7871 [gr-qc]}} (2014).

\bibitem{will2} C. M. Will, \textit{The 1919 measurement of the deflection of light}, Class. Quant. Grav. 32 (2015) no.12, 124001, \href{https://arxiv.org/abs/1409.7812}{{\tt arXiv:1409.7812 [physics.hist-ph]}}.

\bibitem{Ligo1} B. P. Abbott \textit{et al} (LIGO Scientific and Virgo Collaborations), \textit{Observation of Gravitational Waves from a Binary Black Hole Merger}, Phys.Rev.Lett. 116 (2016) no.6, 061102, \href{https://arxiv.org/abs/1602.03837}{{\tt arXiv:1602.03837 [gr-qc]}}.

\bibitem{Ligo2} B. P. Abbott \textit{et al} (LIGO Scientific and Virgo Collaborations), \textit{GW151226: Observation of Gravitational Waves from a 22-Solar-Mass Binary Black Hole Coalescence}, Phys.Rev.Lett. 116 (2016) no.24, 241103, \href{https://arxiv.org/abs/1606.04855}{{\tt arXiv:1606.04855 [gr-qc]}}.

\bibitem{Ligo3} B. P. Abbott \textit{et al} (LIGO Scientific and Virgo Collaborations), \textit{GW170814: A Three-Detector Observation of Gravitational Waves from a Binary Black Hole Coalescence}, Phys.Rev.Lett. 119 (2017) no.14, 141101, \href{https://arxiv.org/abs/1709.09660}{{\tt arXiv:1709.09660 [gr-qc]}}.

\bibitem{Ligo4} B. P. Abbott \textit{et al} (LIGO Scientific and Virgo Collaborations), \textit{GW170817: Observation of Gravitational Waves from a Binary Neutron Star Inspiral}, Phys.Rev.Lett. 119 (2017) no.16, 161101, \href{https://arxiv.org/abs/1710.05832}{{\tt arXiv:1710.05832 [gr-qc]}}.

\bibitem{chan} S. Chandrasekhar, \textit{The mathematical theory of black holes}, Oxford University Press, Nova York (2006).

\bibitem{BCH} J. M. Bardeen, B. Carter, S.W. Hawking, \textit{The Four laws of black hole mechanics}, Commun. Math. Phys. 31 (1973) 161-170.

\bibitem{Grav} C. W. Misner, K. S. Thorne, J. A. Wheller, \textit{Gravitation}, Freeman, San Francisco (1973).

\bibitem{las} M. P. Hobson, G. P. Efstathiou e A. N. Lasenby, \textit{General Relativity - An Introduction for Physicists}, Cambridge University Press, Nova York (2006).


\bibitem{vis1} D. L. Wiltshire, M. Visser, S. M. Scott, \textit{The Kerr spacetime: Rotating black holes in general relativity}, Cambridge University Press, England (2009).

\bibitem{kerr} R. P. Kerr, \textit{Gravitational field of a spinning mass as an example of algebraically special metrics}, Phys.Rev.Lett. 11 (1963) 237-238.

\bibitem{vitor1} M. C. Ferreira, C. F. B. Macedo, V. Cardoso, \textit{Orbital fingerprints of ultralight scalar fields around black holes}, Phys. Rev. D96 (2017) no.8, 083017, \href{https://arxiv.org/abs/1710.00830}{{\tt arXiv:1710.00830 [gr-qc]}}.

\bibitem{Herdeiro1} J. C. Degollado, C. A. R. Herdeiro, \textit{ Stationary scalar configurations around extremal charged black holes}, Gen. Rel. Grav. 45 (2013) 2483-2492, \href{https://arxiv.org/abs/1303.2392}{{\tt arXiv:1303.2392 [gr-qc]}}.

\bibitem{Herdeiro2} C. L. Benone, L. C. B. Crispino, C. Herdeiro, E. Radu, \textit{Kerr-Newman scalar clouds}, Phys. Rev. D90 (2014) no.10, 104024, \href{https://arxiv.org/abs/1409.1593}{{\tt arXiv:1409.1593 [gr-qc]}}.

\bibitem{Herdeiro3} P. V. P. Cunha, C. A. R. Herdeiro, E. Radu, H. F. Runarsson, \textit{Shadows of Kerr black holes with scalar hair}, Phys. Rev. Lett. 115 (2015) no.21, 211102, \href{https://arxiv.org/abs/arXiv:1509.00021}{{\tt arXiv:1509.00021 [gr-qc]}}.

\bibitem{Herdeiro4} P. V. P. Cunha, C. A. R. Herdeiro, E. Radu, H. F. Runarsson, \textit{Shadows of Kerr black holes with and without scalar hair}, Int. J. Mod. Phys. D25 (2016) no.09, 1641021, \href{https://arxiv.org/abs/1605.08293}{{\tt arXiv:1605.08293 [gr-qc]}}.

\bibitem{Bronnikov} K. A. Bronnikov , S. G. Rubin, \textit{Black Holes, Cosmology and Extra Dimensions}, World Scientific, New Jersey (2013).


\bibitem{Hawking1} S. Hawking, R. Penrose, \textit{The Nature of Spacetime}, Princeton University Press, Princeton (1996).

\bibitem{Hawking2} S. Hawking, R. Penrose, \textit{The Singularities of gravitational collapse and cosmology}, Proc. Roy. Soc. Lond. A314 (1970) 529-548.

\bibitem{Hawking3} G. F. R. Ellis, S. Hawking, \textit{ The Cosmic black body radiation and the existence of singularities in our universe}, Astrophys.J. 152 (1968) 25.

\bibitem{Penrose} R. Penrose, \textit{Gravitational collapse and space-time singularities}, Phys. Rev. Lett. 14 (1965) 57-59.

\bibitem{Sakharov} A. D. Sakharov, \textit{The Initial Stage of an Expanding Universe and the Appearance of a Nonuniform Distribution of Matter}, Zh.Eksp.Teor.Fiz. 49 no.1, 345-358, Sov.Phys.JETP 22 (1966) 241.

\bibitem{Gliner} E. Gliner, \textit{Algebraic properties of the energy-momentum tensor and vacuumlike states of matter}, Sov. Phys. JETP 22, 378 (1966).


\bibitem{Bardeen} 
J. M. Bardeen,{\it{ Non-singular general relativistic gravitational collapse}}, in Proceedings of the International Conference GR5, Tbilisi, U.S.S.R. (1968).


\bibitem{Beato1}
E. Ayon-Beato, A. Garcia, {\it{The Bardeen Model as a Nonlinear Magnetic Monopole}}, Phys. Lett. B 493 (2000) 149-152, \href{http://arxiv.org/abs/gr-qc/0009077}{{ \tt gr-qc/0009077}}.

\bibitem{Beato2}
E. Ayón-Beato, A. García, {\it{New Regular Black Hole Solution from Nonlinear Electrodynamics}}, Phys. Lett. B {\bf 464}: 25, (1999), \href{http://arxiv.org/abs/hep-th/9911174}{{\tt  hep-th/9911174}}.

\bibitem{Beato3}
Eloy Ayón-Beato, Alberto García, {\it{Regular Black Hole in General Relativity Coupled to Nonlinear Electrodynamics}}, Phys. Rev. Lett. {\bf 80}: 5056-5059, (1998), \href{http://arxiv.org/abs/gr-qc/9911046}{{\tt   	gr-qc/9911046}}.


\bibitem{Beato4} E. Ayon-Beato, A. Garcia \textit{Four parametric regular black hole solution}, Gen. Rel. Grav. 37 (2005) 635, \href{https://arxiv.org/abs/hep-th/0403229}{{\tt arXiv:hep-th/0403229}}.



\bibitem{Kirill1}
K. A. Bronnikov, {\it{Regular Magnetic Black Holes and Monopoles from Nonlinear Electrodynamics}}, Phys.Rev.D {63}: 044005, (2001), \href{http://arxiv.org/abs/gr-qc/0006014}{{\tt  gr-qc/0006014}}.

\bibitem{Kirill12}
K. A. Bronnikov, {\it{Comment on `Regular black hole in general relativity coupled to nonlinear electrodynamics'}}, Phys.Rev.Lett. 85 (2000) 4641.


\bibitem{Irina}
Irina Dymnikova, {\it{Regular electrically charged structures in Nonlinear Electrodynamics coupled to General Relativity}}, Class.Quant.Grav {\bf .21}: 4417-4429, (2004), \href{http://arxiv.org/abs/gr-qc/0407072}{{\tt  	gr-qc/0407072}}.


\bibitem{Stefano} S. Ansoldi, \textit{Spherical black holes with regular center: A Review of existing models including a recent realization with Gaussian sources}, \href{https://arxiv.org/abs/0802.0330}{{\tt   arXiv:0802.0330 [gr-qc]}}.


\bibitem{Leonardo2}
Leonardo Balart, Elias C. Vagenas, {\it{Regular black holes with a nonlinear electrodynamics source}}, Phys. Rev. D {\bf 90}, 124045 (2014), \href{http://arxiv.org/abs/1408.0306}{{\tt arXiv:1408.0306}}.

\bibitem{balart}
Leonardo Balart, Elias C. Vagenas, {\it{Regular black hole metrics and the weak energy condition}}, Phys.Lett. B {\bf 730} (2014) 14-17, \href{http://arxiv.org/abs/arXiv:1401.2136}{{\tt   arXiv:1401.2136 [gr-qc]}}.


\bibitem{Nami} N. Uchikata, S. Yoshida, T. Futamase, \textit{New Solutions of Charged Regular Black Holes and Their Stability}, Conference: C12-07-01.1, p.1207-1209 Proceedings.

\bibitem{Ponce} J. Ponce de Leon, \textit{Regular Reissner-Nordström black hole solutions from linear electrodynamics}, Phys. Rev. D95 (2017) no.12, 124015, \href{https://arxiv.org/abs/1706.03454}{{\tt   arXiv:1706.03454 [gr-qc]}}.

\bibitem{Manuel1} M. E. Rodrigues, E. L.B. Junior, M. V. de S. Silva, \textit{Using Dominant and Weak Energy Conditions for build New Classes of Regular Black Holes}, \href{https://arxiv.org/abs/1705.05744}{{\tt  arXiv:1705.05744 [physics.gen-ph]}}.

\bibitem{bambi} C. Bambi, L. Modesto, \textit{Rotating regular black holes}, Phys. Lett. B721 (2013) 329-334, \href{https://arxiv.org/abs/1302.6075}{{\tt arXiv:1302.6075 [gr-qc]}}.

\bibitem{neves} J. C. S. Neves, A. Saa, \textit{Regular rotating black holes and the weak energy condition}, Phys. Lett. B734 (2014) 44-48, \href{https://arxiv.org/abs/1402.2694}{{\tt  arXiv:1402.2694 [gr-qc]}}.

\bibitem{toshmatov} B. Toshmatov, B. Ahmedov, A. Abdujabbarov, Z. Stuchlik, \textit{Rotating Regular Black Hole Solution}, Phys. Rev. D89 (2014) no. 10, 104017, \href{https://arxiv.org/abs/1404.6443}{{\tt  arXiv:1404.6443 [gr-qc]}}.

\bibitem{azreg} M. Azreg-Aïnou, \textit{Generating rotating regular black hole solutions without complexification}, Phys. Rev. D90 (2014) no. 6, 064041, \href{https://arxiv.org/abs/1405.2569}{{\tt  arXiv:1405.2569 [gr-qc]}}.

\bibitem{DYM} I. Dymnikova, E. Galaktionov, \textit{Regular rotating electrically charged black holes and solitons in non-linear electrodynamics minimally coupled to gravity}, Class. Quant. Grav. 32 (2015) no. 16, 165015, \href{https://arxiv.org/abs/1510.01353}{{\tt arXiv:1510.01353 [gr-qc]}}.

\bibitem{ramon} R. Torres, F. Fayos, \textit{On regular rotating black holes}, Gen. Rel. Grav. 49 (2017) no. 1, 2, Quant. Grav. 32 (2015) no. 16, 165015, \href{https://arxiv.org/abs/1611.03654}{{\tt arXiv:1611.03654 [gr-qc]}}.

\bibitem{berej} W. Berej, J. Matyjasek, D. Tryniecki, M. Woronowicz, \textit{Regular black holes in quadratic gravity}, Gen. Rel. Grav. 38 (2006) 885-906, \href{https://arxiv.org/abs/hep-th/0606185}{{\tt arXiv:hep-th/0606185}}.

\bibitem{rodrigues1} E. L. B. Junior, M. E. Rodrigues, M. J. S. Houndjo, \textit{Regular black holes in $f(T)$ Gravity through a nonlinear electrodynamics source}, JCAP 1510 (2015) 060, \href{https://arxiv.org/abs/1503.07857}{{\tt arXiv:1503.07857 [gr-qc]}}.

\bibitem{rodrigues2} M. E. Rodrigues, E. L. B. Junior, G. T. Marques, V. T. Zanchin, \textit{Regular black holes in $f(R)$ gravity coupled to nonlinear electrodynamics}, Phys. Rev. D94 (2016) no. 2, 024062, \href{https://arxiv.org/abs/1511.00569}{{\tt arXiv:1511.00569 [gr-qc]}}.

\bibitem{rodrigues3} M. E. Rodrigues, J. C. Fabris, E. L. B. Junior, G. T. Marques, \textit{Generalisation for regular black holes on general relativity to $f(R)$ gravity}, Eur. Phys. J. C76 (2016) no. 5, 250, \href{https://arxiv.org/abs/1601.00471}{{\tt arXiv:1601.00471 [gr-qc]}}.


\end{thebibliography}
\end{document}